\documentclass[a4paper, 11pt]{article}
\usepackage{}
\usepackage{amsfonts}
\usepackage{dsfont}
\usepackage{mathrsfs}
\usepackage{amssymb}
\usepackage{bbm}
\usepackage{epsfig}
\usepackage{amsmath}
\usepackage{appendix}
\usepackage{color}
\usepackage[english]{babel}

\def\Reals{\mathop{\hbox{\mit I\kern-.2em R}}\nolimits}

\def\Complexes{{\hbox{\mit C\kern-.46em
            \vrule depth 0ex height 1.4ex width .05em\kern.41em}}}

\newtheorem{thm}{Theorem}[section]
\newtheorem{defn}{Definition}[section]

\newtheorem{lem}{Lemma}[section]
\newtheorem{remark}{Remark}[section]
\newtheorem{prop}{Proposition}[section]

\title{\bf  Reaching an Optimal Consensus: Dynamical\\ Systems that Compute  Intersections of Convex Sets\footnote{This work has been supported in part
by the Knut and Alice Wallenberg Foundation, the Swedish Research
Council, KTH SRA TNG,  and the NNSF of China under Grant 61174071.}}
\date{}

\author{Guodong Shi, Karl Henrik Johansson\thanks{G. Shi and K. H. Johansson are with ACCESS Linnaeus Centre, School of Electrical Engineering,
Royal Institute of Technology, Stockholm 10044, Sweden.
       Email: {\tt\small guodongs@kth.se, kallej@ee.kth.se}} and Yiguang Hong\thanks{ Y. Hong is with Key Laboratory of Systems and Control, Institute of Systems
       Science, Chinese Academy of Sciences, Beijing 100190, China.
       Email: {\tt\small  yghong@iss.ac.cn}}
}

\begin{document}

\maketitle

\begin{abstract}
In this paper, multi-agent systems minimizing a sum of objective functions, where each component is only known to a particular node,
is considered for continuous-time dynamics with time-varying
interconnection topologies.   Assuming that each node can observe a convex solution set of its optimization component, and the intersection of all such sets is nonempty, the considered optimization problem is converted to an intersection computation problem. By a simple distributed control rule, the considered multi-agent system with continuous-time dynamics achieves not
only a consensus, but also an optimal agreement within the
optimal solution set of the overall optimization objective.   Directed and bidirectional communications are studied, respectively,  and connectivity conditions are given to ensure a global optimal consensus.  In this way,  the corresponding intersection computation problem is solved by the proposed  decentralized continuous-time algorithm. We establish several important properties of the distance functions with respect to the global optimal solution set and a class of invariant sets with the help of convex and non-smooth analysis.
\end{abstract}

{\bf Keywords:} Multi-agent systems, Optimal consensus, Connectivity Conditions, Distributed optimization, Intersection computation

\section{Introduction}
In recent years, multi-agent dynamics has been intensively
investigated in various areas including engineering, natural
science, and social science. Cooperative
control of multi-agent systems is an active research topic, and
rapid developments of distributed control protocols via
interconnected communication have been made to achieve the
collective tasks, e.g.,  \cite{tsi, jad03, saber04,
mor,sabertac, mar, ren, shi09, xiao, lwang}. However, fundamental challenges
still lie in finding suitable tools to describe and design the
dynamical behavior of these systems and thus providing insights in
their functioning principles. Different from the classical control
design, the multi-agent studies aim at fully exploiting, rather than
avoiding, interconnection between agents in analysis and synthesis
in order to deal with distributed design and large-scale information
process.

Consensus is a basic problem of the study of multi-agent
coordination, which usually requires that all the agents  achieve the same state, e.g., a certain relative position or velocity. To achieve collective behavior,
connectivity plays a key role in the coordination of multi-agent
network, and various connectivity conditions have been used to describe frequently
switching topologies in different cases.  The ``joint connection" or
similar concepts are important in the analysis of stability and
convergence to guarantee a suitable convergence. Uniform
joint-connection, i.e., the joint graph is connected during all intervals which are longer than a constant,  has been employed for different consensus problems
\cite{tsi, jad03, lin07, hong07, cheng}. On the other hand,
$[t,\infty)$-joint connectedness, i.e., the joint graph is connected in the time intervals $[t,\infty)$,  is necessary \cite{ shi09, mor}, and therefore the most general form to secure
the global coordination.

Moreover, distributed optimization of a sum of convex objective functions, $\sum_{i=1}^N f_i(z)$, where each component $f_i$ is known only to node $i$,  has attracted much attention in
recent years, due to its wide application in multi-agent systems and wireless networks \cite{rabbat, bj07, ram07, bj08, lu1}.  A class of subgradient-based incremental  when some estimate of the optimal solution  can be passed over the network via deterministic  or randomized iteration were studied in \cite{rabbat, bj07, ram}. Then  a non-gradient-based algorithm was proposed in \cite{lu1}, where each node starts at its own optimal solution and updates using a pairwise equalizing protocol.   In view of multi-agent systems, the local information transmitted over the neighborhood is usually limited to a convex combination of its neighbors \cite{tsi, jad03, mor}. Combining the ideas of consensus algorithms and subgradient methods, a number of significant results were obtained.   A subgradient method in combination with consensus steps was given for solving coupled optimization
problems with fixed undirected topology in
\cite{bj08}. { Then, an important work on multi-agent optimization was \cite{nedic2}, where a decentralized algorithm was proposed as a simple sum of an averaging (consensus) part and a subgradient part, and convergence bounds for a distributed multi-agent model  under various connectivity conditions  were shown.    Constrained consensus and optimization  were further studied in \cite{nedic4}, where each agent was always restricted  in its own convex set. A ``projected consensus algorithm" was presented to solve the constrained consensus problem in which each agent takes averaging and projection steps alternatively, and  it was generalized to ``projected subgradient algorithm" with optimization goal also took into consideration \cite{nedic4}. }

{ Most of the literature on optimization and consensus algorithms is in discrete time, and  it is usually hard for the considered agents to reach both consensus and optimum unless the weights rule of the links, the step size in the iteration and the connectedness of the communication graph are properly selected \cite{nedic2, nedic4, bj07}.  Few researchers  have considered continuous-time  agent dynamics that solves a distributed optimization problem. However, dynamical system solution to optimization problem  is of great interest  since a simple vector-field  solution may provide important geometrical insights.   The classical Arrow-Hurwicz-Uzawa flow was shown to converge to the set of saddle points for a constrained convex optimization problem \cite{ahu}. Then in \cite{brockett}, a simple and elegant continuous-time protocol was presented which solves   linear programming problems.}

{ The goal of this paper is to establish a simple distributed continuous-time control law which can ensure consensus and minimize $\sum_{i=1}^N f_i(z)$ asymptotically. Each optimal solution set, $X_i$ of optimization objective $f_i(z)$, is assumed to be a convex set observed only by node $i$. Assuming that the intersection set, $\bigcap_{i=1}^N X_i$, is nonempty, the optimal solution set of the group objective  becomes this intersection set, and the considered optimization problem is  then converted to a distributed intersection computation problem.  In fact,  computing several convex sets' intersection is a classical problem, and ``alternating projection algorithm" was a standard solution, in which the algorithm is carried out by iteratively projecting onto each set \cite{inter1, inter2, inter3}. The  ``projected consensus algorithm"  presented in \cite{nedic4} can be viewed  as its generalized version.  The intersection computation problem is also of interest in the study of computational geometry, a branch of computer science \cite{cs2, cs3}.} Hence, an important motivation for our work is to provide a system-theoretic insight into the convergence properties of certain distributed optimization problems. Similar to the continuous-time approximation of recursive algorithms \cite{ljung}  and constrained optimizations \cite{ahu,brockett}, we establish a suitable dynamical model for such analysis. Also by itself, the considered continuous-time distributed optimization problem has many applications, e.g., wireless resource allocation \cite{rabbat, bj07}, formation control \cite{mar, shi09, tantac}, and mobile sensing \cite{sabertac,jmf}.

{ In this paper, we present a simple dynamical system solution to this convex intersection computation problem, as the sum of a consensus part and a projection part. Since this projection part can be viewed as a special subgradient information, this protocol is actually a continuous-time version of the algorithm proposed in \cite{nedic2}.
We show that an optimal consensus (i.e., consensus within the global optimal solution set),  can be achieved  under  time-varying communications.  Both directed and bidirectional cases are investigated, and  sharp  connectivity conditions are obtained in the sense that a general optimal consensus will no longer hold for a general model with weaker connectedness. Additionally, we use quite general weights rule which allow the weight of each arc in the communication graph to depend on time or system state. }

The rest of the paper is organized as follows.  In Section 2, some preliminary concepts are introduced. In Section 3, we formulate the considered optimal consensus problem, and the main results are shown.
Then, in Sections 4 and 5, convergence to the optimal solution set  and global consensus  are analyzed, respectively, based on which the proofs of the main results are obtained.  Finally, in Section 6 concluding remarks are
given.

\section{Preliminaries}
In this section, we introduce some  notations and theories on graph theory \cite{god}, convex analysis \cite{boyd, rock} and nonsmooth analysis \cite{rou}.

\vspace{2mm}

A directed graph (digraph) $\mathcal
{G}=(\mathcal {V}, \mathcal {E})$ consists of a finite set
$\mathcal{V}$ of nodes and an arc set
$\mathcal {E}$, in which an arc is an ordered pair of
distinct nodes of $\mathcal {V}$.  An element $(i,j)\in\mathcal {E}$ describes
an arc which leaves $i$ and enters $j$.  A {\it walk} in digraph $\mathcal
{G}$ is an alternating sequence $\mathcal
{W}:
i_{1}e_{1}i_{2}e_{2}\dots e_{m-1}i_{m}$ of nodes $i_{\kappa}$ and
arcs $e_{\kappa}=(i_{\kappa},i_{\kappa+1})\in\mathcal {E}$ for
$\kappa=1,2,\dots,m-1$. A walk  is called a {\it path}
if the nodes of this walk are distinct, and a path from $i$ to
$j$ is denoted as $i\rightarrow j$. $\mathcal
{G}$ is said to be {\it strongly connected} if it contains path $i\rightarrow j$ and $j\rightarrow i$ for every pair of nodes $i$ and $j$. A digraph $\mathcal {G}$ is called to be {\it bidirectional}
when for any two nodes $i$ and $j$, $(i,j)\in\mathcal{E}$  if and only if $(j,i)\in\mathcal{E}$. Ignoring the direction of the arcs, the connectedness of a bidirectional digraph will be transformed to that of the corresponding undirected graph. A time-varying graph is defined as $\mathcal
{G}_{\sigma(t)}=(\mathcal {V},\mathcal {E}_{\sigma(t)})$ with
$\sigma:[0,+\infty)\rightarrow \mathcal {Q}$ as a piecewise constant function,
where $\mathcal {Q}$ is a finite set indicating all possible graphs. Moreover, the joint graph of $\mathcal
{G}_{\sigma(t)}$ in
time interval $[t_1,t_2)$ with $t_1<t_2\leq +\infty$ is denoted as
$\mathcal {G}([t_1,t_2))= \cup_{t\in[t_1,t_2)}
\mathcal {G}(t)=(\mathcal {V},\cup_{t\in[t_1,t_2)}\mathcal
{E}_{\sigma(t)})$.

\vspace{2mm}

A set $K\subset \mathds{R}^m$ is said to be {\it convex} if $(1-\lambda)x+\lambda
y\in K$ whenever $x\in K,y\in K$ and $0\leq\lambda \leq1$.
For any set $S\subset \mathds{R}^m$, the intersection of all convex sets
containing $S$ is called the {\it convex hull} of $S$, denoted by
$co(S)$. The next lemma can be found in \cite{aubin}.

\begin{lem}\label{lems1}
Let $K$ be a  subset of $\mathds{R}^m$. The convex hull $co (K)$ of $K$ is the set of elements of the form
$$
x=\sum_{i=1}^{m+1} \lambda_i x_i,
$$
where $\lambda_i\geq 0, i=1,\dots,m+1$ with $\sum_{i=1}^{m+1} \lambda_i=1$ and $x_i\in K$.
\end{lem}

Let $K$ be a closed convex subset in $\mathds{R}^m$ and denote
$|x|_K\doteq\inf_{y\in K}| x-y |$ as the distance between $x\in \mathds{R}^m$ and $K$, where $|\cdot|$
denotes the Euclidean norm.  There is a unique element ${P}_{K}(x)\in K$ satisfying
$|x-{P}_{K}(x)|=|x|_K$ associated to any
$x\in \mathds{R}^m$ \cite{aubin}.  The map ${P}_{K}$ is called the {\it projector} onto $K$. We also have
\begin{equation}\label{r9}
\langle {P}_{K}(x)-x,{P}_{K}(x)-y\rangle\leq 0,\quad \forall y\in
K.
\end{equation}
Moreover, ${P}_{K}$ has the following non-expansiveness property:
\begin{equation}\label{r8}
|{P}_{K}(x)-{P}_{K}(y)|\leq|x-y|, x,y\in \mathds{R}^m.
\end{equation}
Clearly, $|x|_K^2$ is continuously differentiable at point $x$, and (see \cite{aubin})
\begin{equation}\label{r10}
\nabla |x|_K^2=2(x-{P}_{K}(x)).
\end{equation}

The following lemma was obtained in \cite{shi09}, which is useful in
what follows.

\begin{lem}\label{lem3}
Suppose $K\subset \mathds{R}^m$ is a convex set and
$x_a,x_b\in \mathds{R}^m$. Then
\begin{equation}\label{7}
\langle x_a-{P}_K(x_a),x_b-x_a\rangle\leq
|x_a|_K\cdot     \left.| |x_a|_K-|x_b|_K \right|.
\end{equation}
Particularly, if $|x_a|_K>|x_b|_K$, then
\begin{equation}\label{8}
\langle x_a-{P}_K(x_a),x_b-x_a\rangle\leq -|x_a|_K\cdot(
|x_a|_K-|x_b|_K).
\end{equation}
\end{lem}

\vspace{2mm}

Next, the upper {\it Dini
derivative} of a continuous function $h: (a,b)\to \mathds{R}$ ($-\infty\leq a<b\leq \infty$) at $t$ is defined as
$$
D^+h(t)=\limsup_{s\to 0^+} \frac{h(t+s)-h(t)}{s}.
$$
When $h$ is continuous on $(a,b)$, $h$ is
non-increasing on $(a,b)$ if and only if $ D^+h(t)\leq 0$ for any
$t\in (a,b)$. The next
result is given for the calculation of Dini derivative
(see \cite{dan,lin07}).

\begin{lem}
\label{lem1}  Let $V_i(t,x): \mathds{R}\times \mathds{R}^d \to \mathds{R}\;(i=1,\dots,n)$ be
$C^1$ and $V(t,x)=\max_{i=1,\dots,n}V_i(t,x)$. If $
\mathcal{I}(t)=\{i\in \{1,2,\dots,n\}\,:\,V(t,x(t))=V_i(t,x(t))\}$
is the set of indices where the maximum is reached at $t$, then
$
D^+V(t,x(t))=\max_{i\in\mathcal{ I}(t)}\dot{V}_i(t,x(t)).
$
\end{lem}

\vspace{2mm}
Finally, consider a system
\begin{equation}
\label{i1} \dot{x}=f(t,x),
\end{equation}
where $f:\mathds{R}\times \mathds{R}^d\rightarrow \mathds{R}^d$ is piecewise continuous in $t$
and continuous in $x$. Let $x(t)=x(t,t_0,x^0)$ be a solution of
(\ref{i1}) with initial condition $x(t_0)=x^0$. Then $\Omega_0\subset \mathds{R}^d$ is called a {\it positively invariant
set} of (\ref{i1}) if, for any $t_0\in R$ and any $x^0\in\Omega_0$,
$x(t,t_0,x^0)\in\Omega_0$ when $t\geq t_0$.

\section{Problem Formulation and Main Results}
In this section, we first define the considered optimal consensus problem. We propose a multi-agent optimization model and  a distributed control law to solve this optimization problem. Then the main results are presented on connectivity conditions which can ensure an optimal consensus globally.

\subsection{Multi-agent Model}

Consider a multi-agent system with agent set $\mathcal
{V}=\{1,2,\dots,N\}$, for which the dynamics of each agent is a first-order integrator:
\begin{equation}\label{2}
\dot{x}_i=u_i, \quad i=1,\dots,N
\end{equation}
where $x_i\in \mathds{R}^m$ represents the state of agent $i$, and $u_i$ is
the control input.

The communication in the multi-agent network is modeled as a time-varying graph $\mathcal {G}_{\sigma(t)}=(\mathcal {V},\mathcal
{E}_{\sigma(t)})$. Moreover, node $j$ is said to be a {\it neighbor} of $i$ at time $t$ when there is an arc $(j, i)\in \mathcal
{E}_{\sigma(t)}$, and $N_i(\sigma(t))$ represents the set of agent $i$'s neighbors at time $t$. As usual in the literature \cite{jad03,lin07,shi09}, an assumption is given to the variation of $\mathcal {G}_{\sigma(t)}$.

\noindent {\bf A1} {\it (Dwell Time)} There is a lower bound constant $\tau_D>0$ between two consecutive
switching time instants of $\sigma(t)$.

We have the following definition.
\begin{defn}
(i) $\mathcal
{G}_{\sigma(t)}$ is said to be {\it uniformly jointly strongly connected} (UJSC) if there exists a constant $T>0$ such that $\mathcal {G}([t,t+T))$ is strongly connected for any $t\geq0$.

(ii) Assume that $\mathcal {G}_{\sigma(t)}, t\geq 0$ is bidirectional. $\mathcal
{G}_{\sigma(t)}$ is said to be {\it infinitely jointly connected} (IJC)  if $\mathcal
{G}([t,+\infty))$ is connected for all $t\geq0$.
\end{defn}
\begin{remark}
$[t,+\infty)$-joint connectedness for all $t\geq0$ is equivalent to that there exists an unbounded time sequence
$
0\leq t_1<\dots<t_k<t_{k+1}<\dots$
such that $\mathcal
{G}([t_k,t_{k+1}))$ is connected for all $k=1,2,\dots$. Note that it does not require an upper bound for $|t_{k+1}-t_k|$ in the definition.
\end{remark}
The objective for this group of
autonomous agents is to reach a consensus, and meanwhile to cooperatively solve the
following optimization problem
\begin{equation}\label{1}
\min_{z\in \mathds{R}^m}\ \ \ \sum_{i=1}^N f_i(z)
\end{equation}
where $f_i:\mathds{R}^m\rightarrow \mathds{R}$ represents the cost function of agent
$i$, observed by agent $i$ only, and $z$ is a decision vector. We suppose the optimal solution set of
each component $f_i$ exists,  denoted $
 X _i\doteq \{v\ | f_i(v)=\min\limits_{z\in \mathds{R}^m} f_i(z)\}$.

We impose the following assumptions.

\noindent{\bf A2} {\it (Convexity)} $X_1,\dots,X_N$, are closed convex sets.

\noindent{\bf A3} {\it (Nonempty Intersection)}  $X_0\doteq\bigcap\limits_{i=1}^{N}  X _i$  is nonempty and bounded.

\begin{remark}
The assumption that each $X_i$ is a convex set is quite general, and it is not hard to see that this assumption will be satisfied  as long as each $f_i$ is a convex function. Moreover,  since the
intersection of convex sets is a convex set itself, $X_0$ is a convex set with the convexity of each $X_i$. Additionally, with A3, it is obvious to see that $ X _0$ is compact, and it is the optimal solution set of  (\ref{1}).
\end{remark}

\subsection{Distributed Control}

Denote $x=(x_1^T,\dots,x_N^T)^T\in \mathds{R}^{mN}$ and let the continuous function
$a_{ij}(x,t)>0$ be the weight of arc $(j,i)$, for $i,j\in
\mathcal {V}$.  Then we present the following distributed control law:
\begin{equation}\label{6}
u_{i}=\sum\limits_{j \in
N_i(\sigma(t))}a_{ij}(x,t)(x_j-x_i)+  P _{ X _i}(x_i)-x_i, \;i=1,\dots,N
\end{equation}

\begin{remark}
 We write the arc weight
$a_{ij}(x,t)$ in a quite general form showing that this weight function can be time-varying and may depend nonlinearly on the state. Note that this doesn't mean global information is required for the control design.
\end{remark}

\begin{remark}
When  $X _i$ can be observed by node $i$, ${P}_{ X _i}(x_i(t))-x_i(t)$ can be easily obtained. For instance, node $i$ may first establish a local coordinate system, and then construct a function $h(z)=|z|^2_{X_i}$ to compute $\nabla h(z)$ within this coordinate system. Then by (\ref{r10}), we have ${P}_{ X _i}(x_i(t))-x_i(t)=-1/2\nabla h(z)|_{z=x_i(t)}$.
\end{remark}


Another assumption is made on  each $a_{ij}(x,t),i,j=1,2,...,N$.

\noindent{\bf A4} {\it (Weights Rule)} There are $a^\ast>0$ and
$a_\ast>0$ such that
$$
 a_\ast\leq a_{ij}(x, t)\leq a^\ast,\quad
 t\in \mathds{R}^+,x\in \mathds{R}^{mN}.
$$

In this paper, we assume that Assumptions A1-A4 always hold. With (\ref{2}) and (\ref{6}), the closed loop system is expressed by
\begin{equation}\label{9}
\dot{x}_{i}=\sum\limits_{j \in
N_i(\sigma(t))}a_{ij}(x,t)(x_j-x_i)+  P _{ X _i}(x_i)-x_i, \;i=1,\dots,N.
\end{equation}

\begin{remark}
By the non-expansiveness property (\ref{r8}), the convex projection $P _{ K}(z)$ is  continuous for all $z\in\mathds{R}^m$ for any closed convex set $K\subseteq R^m$. Therefore, a solution of (\ref{9}) exists at least over a finite interval for any initial condition $x(t_0)$. Note that the solution is not necessarily  unique. As will be shown in Remark 4.1, it also exists in $[t_0,+\infty)$.
\end{remark}

\begin{remark}
 Since the projection term can be viewed as a subgradient  for the special case $f_i(z)=|z|_{X_i}^2/2$, (\ref{9}) is actually a continuous-time version of the algorithm proposed in \cite{nedic2}, which has the form of  the sum of a consensus term and a subgradient term. On the other hand, in \cite{nedic4}, a ``projected consensus algorithm" was presented to solve the same intersection computation problem in which each agent takes consensus and projection steps alternatively. Note that there is some essential difference between  (\ref{9}) and the ``projected consensus algorithm" in \cite{nedic4}, because (\ref{9}) takes advantage of the consensus and projection information at the same time instant. It is not hard to construct  examples in which each node $i$ would never enter its own set $X
 _i$ along the trajectories of (\ref{9}).
\end{remark}

\begin{figure}
\centerline{\epsfig{figure=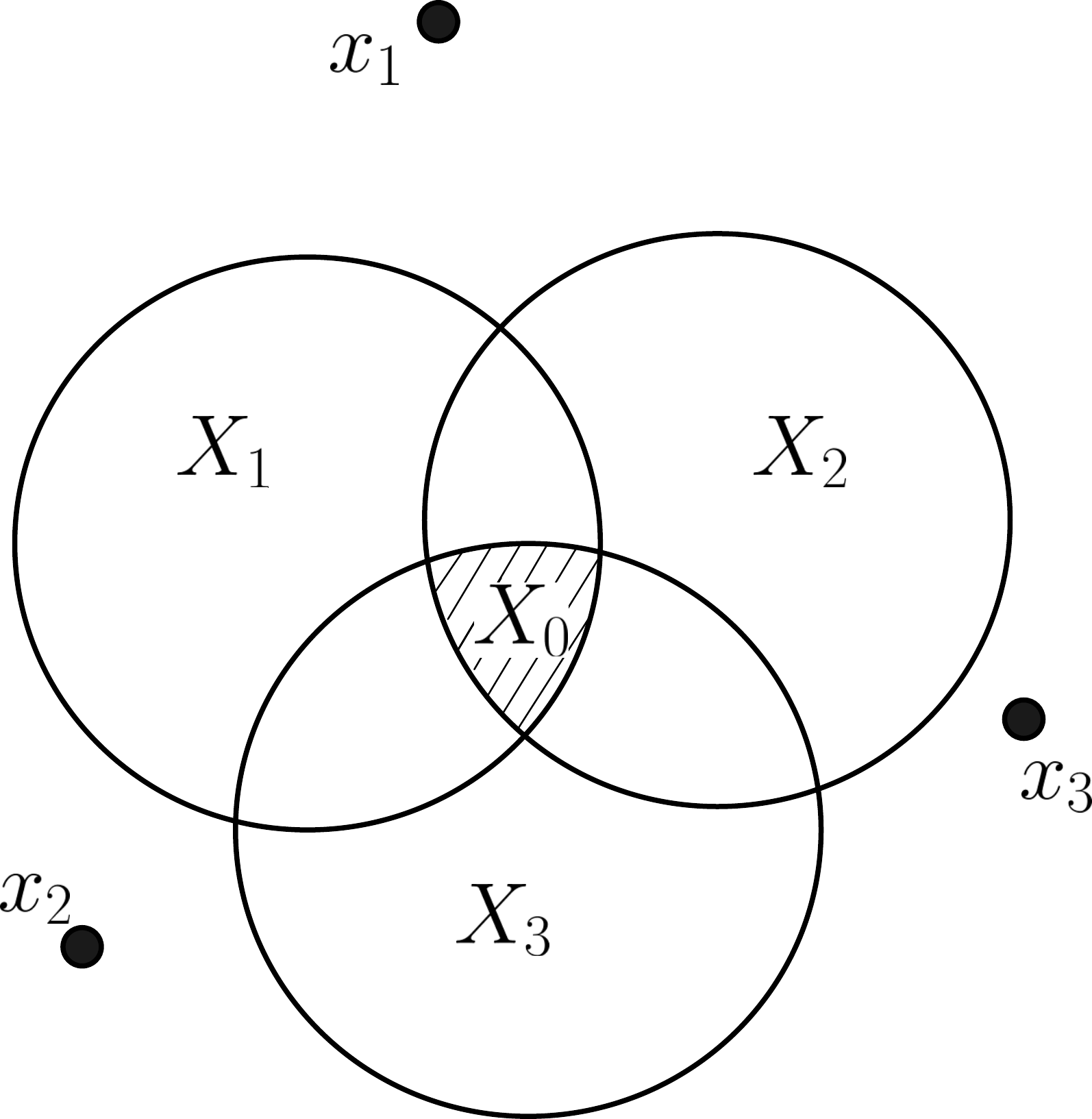, width=0.4\linewidth=0.2}}
\caption{The goal of the agents is to achieve a consensus in $ X _0$.}\label{fig0}
\end{figure}

 Let $x(t)$ be the trajectory of (\ref{9}) with initial condition $x^0=x(t_0)=(x_1^T(t_0),\dots,x_N^T(t_0))^T\in \mathds{R}^{mN}$. Then the considered optimal consensus is defined as following (see Fig. \ref{fig0}).

\begin{defn}
(i) A  global {\it optimal set convergence} of (\ref{9}) is achieved if for all $x^0\in \mathds{R}^{mN}$, we have
\begin{equation}\label{3}
\lim_{t\rightarrow +\infty} |x_i(t)|_{ X _0}=0,\quad i=1,\dots,N.
\end{equation}

(ii) A global {\it consensus} of (\ref{9}) is achieved if for all $x^0\in \mathds{R}^{mN}$, we have
\begin{equation}\label{4}
\lim_{t\rightarrow +\infty} |x_i(t)-x_j(t)|=0,\quad i,j=1,\dots,N.
\end{equation}

(iii) A  global {\it optimal consensus} is achieved of (\ref{9}) if both (i) and (ii) hold.
\end{defn}

\begin{remark}It is easy to find that, based on the analysis methods we provide,  all the results obtained in this paper will still hold if the control law (\ref{9}) is replaced by $$
\dot{x}_{i}=\sum\limits_{j \in
N_i(\sigma(t))}a_{ij}(x,t)(x_j-x_i)+ b_i(x_i,t)( P _{ X _i}(x_i)-x_i), \;i=1,\dots,N$$
for some scalar functions $0<b_\ast\leq b_i(x_i,t), i=1,\cdots,N$ with $b_\ast>0$ being a constant. Here we just choose the form of (\ref{9}) to make the statements and proofs simplified.
\end{remark}

\subsection{Main Results}

In this subsection, we present the main results  on optimal consensus.

First the following conclusion is our main result for directed graphs.

\begin{thm}\label{thm4}
System (\ref{9}) achieves a global optimal consensus if   $\mathcal
{G}_{\sigma(t)}$ is UJSC.
\end{thm}


We say the communications over the considered multi-agent network are bidirectional if $\mathcal {G}_{\sigma(t)}$ is a bidirectional graph for all $t\geq t_0$. Note that, this does not imply that the arc weights, $a_{ij}(x,t), i, j=1,\dots,N$, are symmetric.  Then we have the following main result on optimal consensus for the bidirectional case.

\begin{thm}\label{thm1} System (\ref{9}) with bidirectional communications achieves a global optimal consensus if (and in general only if) $\mathcal
{G}_{\sigma(t)}$ is IJC.
\end{thm}

Theorem \ref{thm1} shows that the connectedness conditions to reach an optimal consensus can be relaxed for bidirectional communications without requiring a uniform bound of the length of intervals in the definition of connectivities.




\begin{remark}
Let us explain what ``in general only if" means in  Theorems 3.2.
Clearly, the connectivity condition proposed in Theorem 3.2 is not a necessary condition to ensure a global optimal consensus for a particular optimization problem (\ref{1}). However, in regard to a global optimal consensus for all possibilities of $X_1,\dots,X_N$, simple examples could show that this IJC assumption is also necessary using the same idea studying state agreement problem  in \cite{mor,shi09}. In fact, as long as $\bigcap_{i=1}^N X_i$ is not a singleton, it can be easily shown that consensus cannot be guaranteed for all initial conditions. Therefore,  from this perspective, Theorem \ref{thm1} gives ``sharp" connectivity conditions for  a global optimal consensus of system (\ref{9}).
\end{remark}


\begin{remark}
If A3, the nonempty intersection assumption, is removed, control law (\ref{9}) becomes a special case of the target aggregation controller studied in \cite{shi09} with respect to $co(\bigcup_{i=1}^N X_i)$. In this case, under proper connectivity assumptions (even each node cannot always obtain the information of $X_i$), it can be shown that (\ref{9}) will lead the network to converge into $co(\bigcup_{i=1}^N X_i)$ \cite{shi09}. The dynamics within $co(\bigcup_{i=1}^N X_i)$ can be complicated, and the optimal consensus will fail since there is no longer a simple expression of  $X_\ast$, the real optimal solution set of (\ref{1}). { However, we guess that in this case the control law (\ref{9}) still implies a suboptimal convergence such that there will be a constant $B$, which does not depend on the initial condition, satisfying $\limsup_{t\rightarrow\infty}|x_i(t)|_{X_\ast}\leq B$ under UJSC connectivity conditions.}
\end{remark}

In order to prove Theorems \ref{thm4} and \ref{thm1},  on one hand, we have to prove all the
agents converge to the global optimal solution set, i.e., $X_0$;
and, on the other hand, we have to verify that a consensus is also
achieved. In fact, the convergence analysis is quite challenging, due to the nonlinearity nature of each weight function $a_{ij}(x,t)$ and the convex projection part in the control law.  In the following two sections, we will focus on the optimal solution set convergence and the consensus analysis, respectively, by which complete the proofs for Theorems \ref{thm4} and \ref{thm1}.

\section{Optimal Set Convergence }

In this section, we prove the optimal solution set convergence for system (\ref{9}). We first establish a method to analyze the distance between the agents and the global optimal set with the help of convex analysis, and then the convergence to $X_0$ for all the agents is proposed under directed and bidirectional communications, respectively.

\subsection{Distance Function}

Define $d_i(t)=|x_i(t)|_{ X _0}^2$ and let
$$
{d}(t)=\max_{i\in\mathcal {V}}d_i(t)
$$
be the maximum among all the agents.  Although ${d}(t)$ may not be continuously differentiable, it is still continuous. Thus, we can analyze the Dini derivative of ${d}(t)$ to study its convergence property. Moreover, it is easy to see that ${d}(t)$ is locally Lipschitz. Then the Dini derivative of ${d}(t)$ is finite for any $t$.

We prove several elementary lemmas  for the following analysis. At first, the following lemma indicates that ${d}(t)$ is nonincreasing.

\begin{lem}
\label{lem2}$D^+{d}(t)\leq 0$ for all $t\geq0$.
\end{lem}
\noindent {\it Proof.} According to (\ref{r10}), one has
\begin{align}\label{14}
\frac{d }{dt}d_i(t) &= 2\langle x_i- P _{ X _0}(x_i), \dot{x}_i\rangle\nonumber\\
&= 2\langle x_i- P _{ X _0}(x_i),
\sum_{j \in N_i(\sigma(t))}a_{ij}(x,t)(x_j-x_i)+ P _{ X _i}(x_i)-x_i\rangle.
\end{align}
Then, based on Lemma \ref{lem1} and denoting $\mathcal{I}(t)$ as the set containing all the agents
that reach the maximum in the definition of ${d}(t)$ at
time $t$, we obtain
\begin{align}\label{11}
D^+{d}(t)&=\max_{i \in \mathcal{I}(t)} \frac{d}{dt}
d_i(t)\nonumber\\
&=2\max_{i \in \mathcal{I}(t)} [\langle x_i- P _{ X _0}(x_i),
\sum_{j \in N_i(\sigma(t))}a_{ij}(x_j-x_i)+ P _{ X _i}(x_i)-x_i\rangle].
\end{align}

Furthermore, for any $i \in \mathcal{I}(t)$, according to (\ref{8}) of Lemma \ref{lem3}, one has
\begin{equation}\label{13}
\langle x_i- P _{ X _0}(x_i),x_j-x_i\rangle\leq 0
\end{equation}
for any $j\in L_i({\sigma(t)})$ since it always holds that $|x_j|_{ X _0}\leq |x_i|_{ X _0}$.

Moreover, in light of (\ref{r9}), we obtain
\begin{equation}
\langle  P _{ X _i}(x_i)- P _{ X _0}(x_i), P _{ X _i}(x_i)-x_i\rangle\leq 0
\end{equation}
since we always have $ P _{ X _0}(x_i)\in  X _i$ for all $i=1,\dots, N$. Therefore, it is easy to see that  for any $i\in \mathcal {V}$,
\begin{align}\label{12}
\langle x_i- P _{ X _0}(x_i), P _{ X _i}(x_i)-x_i\rangle
\leq \langle x_i- P _{ X _i}(x_i),  P _{ X _i}(x_i)-x_i \rangle
= -|x_i|_{ X _i}^2.
\end{align}

Thus, with (\ref{11}), (\ref{13}) and (\ref{12}), one has
\begin{equation}
D^+{d}(t)\leq 2\max_{i \in \mathcal{I}(t)} [-|x_i|_{ X _i}^2]\leq 0.
\end{equation}
Then the proof is completed. \hfill$\square$

\begin{remark}\label{rem1}
According to Lemma \ref{lem2}, $ \{y|\ |y|^2_{ X _0}\leq {d}(t_0)\}$ is a positively invariant set for system (\ref{9}). Since $X_0$ is compact, $ \{y|\ |y|^2_{ X _0}\leq {d}(t_0)\}$ is also compact. This leads to that each solution of (\ref{9}) exists in $[t_0,+\infty)$. Moreover,  if the weight functions $a_{ij}, i,j=1,\dots,N$, are only state-dependent, the continuity implies that there will be $a^\ast\geq a_\ast>0$ such that
\begin{equation}
 a_\ast\leq a_{ij}(x(t))\leq a^\ast,\quad\forall t>0,\;
i,j=1,2,...N
\end{equation}
along trajectory $x(t)$ of system (\ref{9}). In this case, A4 follows automatically, and then needs not to be assumed.
\end{remark}

With Lemma \ref{lem2}, for any initial condition, there exists a constant ${d}^\ast\geq 0$ such that $\lim_{t\rightarrow\infty}{d}(t)={d}^\ast$. Clearly, the optimal solution set convergence will be achieved for system (\ref{9}) if and only if ${d}^\ast=0$. Furthermore, since it always holds that $d_i(t)\leq {d}(t)$, there exist constants $0\leq\theta_i\leq\eta_i\leq {d}^\ast, i=1,\dots, N$ such that
$$
\liminf_{t\rightarrow \infty} d_i(t)=\theta_i, \quad \limsup_{t\rightarrow \infty}d_i(t)= \eta_i.
$$

To establish the optimal set convergence, we also need the following lemmas, whose proofs can be found in the Appendix.
\begin{lem}\label{lem4}
Assume that $\theta_i=\eta_i={d}^\ast, i=1,\dots, N$. Then we have
$\lim_{t\rightarrow+\infty}|x_i(t)|_{ X _i}=0$
for all $i=1,\dots,N$.
\end{lem}
\begin{lem}\label{lem10}
 Assume that either $\mathcal
{G}_{\sigma (t)}$ being UJSC or $\mathcal
{G}_{\sigma (t)}$  being  IJC with bidirectional communications.  Then $\theta_{i}=\eta_{i}={d}^\ast$ for all $i=1,2,\dots,N$.
\end{lem}

\begin{remark} If the network communication graph is undirected, i.e., $i\in N_{j}(\sigma(t))$ if and only if $j\in N_{i}(\sigma(t))$ with $a_{ij}(x,t)\equiv a_{ji}(x,t), i,j=1,\dots,N$, then according to (\ref{14}) and (\ref{12}), we have
\begin{align}\label{38}
\frac{d}{dt} \sum_{i=1}^{N} d_i(t)
&\leq 2\sum_{i=1}^{N}  \sum_{j \in N_i(\sigma(t))}a_{ij}(x,t)\langle x_i- P _{ X _0}(x_i),
x_j-x_i\rangle-2\sum_{i=1}^{N}|x_i|_{ X _i}^2\nonumber\\
&= \sum_{i=1}^{N}  \sum_{j \in N_i(\sigma(t))}a_{ij}(x,t)\langle x_i- P _{ X _0}(x_i),
x_j-x_i\rangle\nonumber\\
&+ \sum_{j=1}^{N}  \sum_{i \in N_j(\sigma(t))}a_{ji}(x,t)\langle x_j- P _{ X _0}(x_j),
x_i-x_j\rangle-2\sum_{i=1}^{N}|x_i|_{ X _i}^2\nonumber\\
&= \sum_{i=1}^{N}  \sum_{j \in N_i(\sigma(t))}a_{ij}(x,t)\langle x_i-x_j+ P _{ X _0}(x_j)- P _{ X _0}(x_i),x_j-x_i\rangle-2\sum_{i=1}^{N}|x_i|_{ X _i}^2.\nonumber
\end{align}
Furthermore, based on (\ref{r9}) and (\ref{r8}), we obtain
\begin{equation}
\langle x_i-x_j+ P _{ X _0}(x_j)- P _{ X _0}(x_i),x_j-x_i\rangle\leq-|x_i-x_j|^2+|x_i-x_j|\cdot| P _{ X _0}(x_j)- P _{ X _0}(x_i)|\leq0\nonumber
\end{equation}
for all $i,j=1,\dots,N$. Therefore,  we have
$$
 \frac{d}{dt} \sum_{i=1}^{N} d_i(t)\leq -2\sum_{i=1}^{N}|x_i(t)|_{ X _i}^2,
$$
which implies
\begin{equation}\label{39}
\sum_{i=1}^{N} \int_0^\infty |x_i(t)|_{ X _i}^2<\frac{N}{2}{d}(t_0)
\end{equation}
immediately based on Lemma \ref{lem2}.

As a result, with (\ref{39}), we can apply Barbalat's lemma on $|x_i(t)|_{ X _i}^2$, and then it follows immediately that
$\lim_{t\rightarrow+\infty}|x_i(t)|_{ X _i}=0, \,i=1,\dots,N$ without the assumptions  of Lemma \ref{lem4}.
\end{remark}

\begin{remark}
Note that, Lemmas \ref{lem2} and \ref{lem4} hold without requiring any connectivity of the system communication graph.
\end{remark}

\subsection{Directed Graphs}
The following conclusion is for optimal set convergence with directed communications.
\begin{prop}\label{thm5}
System (\ref{9}) achieves the global optimal solution set convergence if $\mathcal
{G}_{\sigma(t)}$ is UJSC.
\end{prop}
\noindent {\it Proof.}  According to Lemmas \ref{lem4} and \ref{lem10}, we have $\lim_{t\rightarrow \infty}|x_i(t)|_{ X _{i}}=0$, $i=1,\dots,N$. As a result, for any $\varepsilon>0$, there exists $T_1(\varepsilon)>0$ such that when $t \geq T_1$,
\begin{equation}\label{67}
 |x_i(t)|_{ X _i}\leq\varepsilon, \quad i=1,\dots, N.
\end{equation}

Take ${t}_1=T_1$ and $k_0\in \mathcal {V}$. Defining $h_{k_0}(t)\doteq \max_{i\in\mathcal{V}}|x_i(t)|_{X_{k_0}}$, similarly to the analysis of (\ref{11}), we have that for all $t$,
$$
\frac{d}{dt} h_{k_0}^2(t)\leq 2h_{k_0}(t)\cdot\max_{i=1,\dots,N}|x_i(t)|_{X_i},
$$
which implies $D^+ h_{k_0}(t)\leq \varepsilon, t\geq t_1$. Thus, $h_{k_0}(t)\leq h_{k_0}(t_1)+(N-1)T_0\varepsilon, t\in[{t}_{1},{t}_{1}+(N-1)T_0]$.

 Since $\mathcal
{G}_{\sigma(t)}$ is UJSC, we can find a node $k_1$ such that  $(k_0,k_1)\in\mathcal
{E}_{\sigma(t)}$ for $t\in[\tilde{t}_1,\tilde{t}_1+\tau_D)\subseteq [{t}_{1},{t}_{1}+T_0)$, where  $T_0=T+2\tau_D$. In light of Lemma \ref{lem3} and (\ref{67}),  we have
\begin{align}\label{101}
\frac{d }{dt}|x_{k_1}(t)|^2_{ X _{k_0}}
&=2a_{{k_1}k_0}(x,t)\langle x_{k_1}- P _{ X _{k_0}}(x_{k_1}),
x_{k_0}-x_{k_1}\rangle +2\langle x_{k_1}- P _{ X _{k_0}}(x_{k_1}),\nonumber\\
&\ \ \ \ \ \ \ \ \sum_{j \in N_{k_1}(\sigma(t))\setminus k_0}a_{{k_1}j}(x_j-x_{k_1})
+ P _{ X _{k_1}}(x_{k_1})-x_{k_1}\rangle\nonumber\\
 &\leq-2a_\ast |x_{k_1}(t)|_{ X _{k_0}}(|x_{k_1}(t)|_{ X _{k_0}}-\varepsilon)+2(N-2)a^\ast|x_{k_1}(t)|_{ X _{k_0}}\cdot(h_{k_0}(t_1)\nonumber\\
  &\ \ \ \ \ \ \ \ +(N-1)T_0\varepsilon-|x_{k_1}(t)|_{ X _{k_0}})
+2|x_{k_1}(t)|_{ X _{k_0}}\cdot\varepsilon, \quad t\in[\tilde{t}_1,\tilde{t}_1+\tau_D),
\end{align}
from which we obtain that  for any $t\in[\tilde{t}_1,\tilde{t}_1+\tau_D)$,
\begin{equation}
D^+|x_{k_1}(t)|_{ X _{k_0}}
\leq-(a_\ast+(N-2)a^\ast)|x_{k_1}(t)|_{ X _{k_0}}+(N-2)a^\ast[ h_{k_0}(t_1)+(N-1)T_0\varepsilon]+(1+a_\ast)\varepsilon. \; \nonumber
\end{equation}
Therefore, noticing that $|x_{k_1}(\tilde{t}_1)|_{ X _{k_0}}\leq h_{k_0}(t_1)+(N-1)T_0\varepsilon$ and denoting $\nu_0=e^{-(a_\ast+(N-2)a^\ast)\tau_D}$, one has
\begin{align}
|x_{k_1}(\tilde{t}_1+\tau_D)|_{ X _{k_0}}
&\leq \nu_0|x_{k_1}(\tilde{t}_1)|_{ X _{k_0}}+(1-\nu_0) \cdot\frac{(N-2)a^\ast[ h_{k_0}(t_1)+(N-1)T_0\varepsilon]+(1+a_\ast)\varepsilon}{a_\ast+(N-2)a^\ast}\nonumber\\
&\leq w_0 h_{k_0}(t_1)+M_0\varepsilon,
\end{align}
where \begin{equation}\label{103}
w_0=\frac{(N-2)a^\ast+a_\ast\nu_0}{(N-2)a^\ast+a_\ast}; \quad M_0=\frac{1+[2(N-2)(N-1)T_0+1]a^\ast }{a_\ast+(N-2)a^\ast}.
\end{equation} Proceeding the estimation in time interval $[\tilde{t}_1+\tau_D,{t}_{1}+(N-1)T_0]$ will lead to
\begin{equation}
D^+|x_{k_1}(t)|_{ X _{k_0}}
\leq-(N-1)a^\ast|x_{k_1}(t)|_{ X _{k_0}}+(N-1)a^\ast[ h_{k_0}(t_1)+(N-1)T_0\varepsilon]+\varepsilon, \; t\in [\tilde{t}_1+\tau_D,{t}_{1}+(N-1)T_0],\nonumber
\end{equation}
for all $t\in[\tilde{t}_1+\tau_D,{t}_{1}+(N-1)T_0]$. This  implies
\begin{align}\label{52}
|x_{k_1}(t)|_{X_{k_0}}
&\leq  \varsigma_0(w_0 h_{k_0}(t_1)+M_0\varepsilon)+(1-\varsigma_0)(w_0 h_{k_0}(t_1)+(N-1)T_0\varepsilon+\frac{\varepsilon}{(N-1)a^\ast})\nonumber\\
&\leq w_0 h_{k_0}(t_1)+\tilde{M}_0\varepsilon, \ \ \ \ t\in[\tilde{t}_1+\tau_D,{t}_{1}+(N-1)T_0]
\end{align}
where  \begin{equation}
\varsigma_0=e^{-(N-1)^2a^\ast T_0};\quad \tilde{M}_0=\frac{2+[3(N-1)^2T_0+1]a^\ast}{a_\ast+(N-2)a^\ast}.
\end{equation}

Further, continuing the analysis on time interval $[{t}_{1}+T_0,{t}_{1}+2T_0]$, $k_2$ can be found with a neighbor in $\{k_0,k_1\}$ during $[\tilde{t}_2,\tilde{t}_2+\tau_D)\subseteq [{t}_{1}+T_0,{t}_{1}+2T_0]$. An upper bound for $|x_{k_2}(t)|_{X_{k_0}}$ can be similarly obtained as
 \begin{equation}
|x_{k_2}(t)|_{X_{k_0}}
\leq w_1 h_{k_0}(t_1)+2\tilde{M}_0\varepsilon, \; \; t\in[\tilde{t}_2+\tau_D,{t}_{1}+(N-1)T_0]
\end{equation}
 where $w_1=\frac{(N-2)a^\ast+a_\ast\nu_0^2}{(N-2)a^\ast+a_\ast}$.

 Next, respectively, we repeat the analysis on time intervals $[{t}_{1}+2T_0,{t}_{1}+3T_0],\dots, [{t}_{1}+(N-2)T_0,{t}_{1}+(N-1)T_0]$ for $k_3, \dots, k_{N-1}\in\mathcal {V}$, and we finally reach
\begin{equation}\label{56}
|x_{i}({t}_{1}+(N-1)T_0)|_{X_{k_0}}
\leq w_{N-1}h_{k_0}(t_1)+(N-1)\tilde{M}_0\varepsilon, \; i=1,\dots,N,
\end{equation}
which implies
\begin{equation}
h_{k_0}(t_2)
\leq w_{N-1}h_{k_0}(t_1)+(N-1)\tilde{M}_0\varepsilon,
\end{equation}
where  $t_2=t_1+(N-1)T_0$ and $0<w_{N-1}=\frac{(N-2)a^\ast+a_\ast\nu_0^{N}}{(N-2)a^\ast+a_\ast}<1$.

Denoting $w_{\ast}=w_{N-1}$ and  ${t}_{n+1}={t}_{n}+(N-1)T_0$ for $n=2, \dots$, and by the same analysis on time intervals $[{t}_{n}, {t}_{n+1}], n=2,\dots$, one has
\begin{align}\label{59}
h_{k_0}(t_n)
&\leq w_{\ast}^{n-1}h_{k_0}(t_1)+\sum_{j=1}^{n-1}w_{\ast}^{j-1}(N-1)\tilde{M}_0\varepsilon\nonumber\\
&\leq w_{\ast}^{n-1}h_{k_0}(t_1)+\frac{(N-1)\tilde{M}_0}{1-w_{\ast}}\cdot\varepsilon
\end{align}

 Since $\varepsilon$ in (\ref{59}) can be arbitrarily small, we see that $\lim_{t\rightarrow \infty}|x_i(t)|_{ X _{k_0}}=0$  for all $i,k_0=1,\dots,N$, which immediately implies $\lim_{t\rightarrow \infty}|x_i(t)|_{ X _{0}}=0$. The proof is completed.\hfill$\square$

\subsection{Bidirectional Graphs}
The following conclusion is for optimal set convergence under bidirectional graphs.
\begin{prop}\label{thm2}
System (\ref{9}) achieves the optimal solution set convergence with bidirectional communications if $\mathcal
{G}_{\sigma(t)}$ is IJC.
\end{prop}
\noindent {\it Proof.} Suppose ${d}^\ast>0$. According to Lemmas \ref{lem4} and \ref{lem10}, we have that for all $i=1,\dots,N$,
\begin{equation}\label{s6}
\lim_{t\rightarrow\infty} |x_i(t)|_{X_0}=\sqrt{d^\ast}, \quad \lim_{t\rightarrow\infty} |x_i(t)|_{X_i}=0.\;
\end{equation}
This implies, for any $\varepsilon>0$, we have that $x_i(t)\in \mathcal{B}_0(\varepsilon)\cap \mathcal{B}_i(\varepsilon)$ for sufficiently large $t$, where $\mathcal{B}_0(\varepsilon)\doteq \{y|\sqrt{{d}^\ast+\varepsilon}\leq|y|_{X_0}\leq \sqrt{{d}^\ast+\varepsilon}\}$ and $\mathcal{B}_i(\varepsilon)\doteq \{y||y|_{X_i}\leq \varepsilon\}, i=1,\dots, N$. Then we see from (\ref{14}) that the derivative of $ |x_i(t)|^2_{X_0}$ is globally Lipschitz. Therefore,  based on Barbalat's lemma, we know
\begin{equation}\label{s5}
\lim_{t\rightarrow\infty}\frac{d}{dt} |x_i(t)|^2_{X_0}=0.
\end{equation}

Define $\mathcal{E}_\infty\doteq\{(i,j)|(i,j)\in \mathcal{E}_{\sigma(t)}\ \mbox{for infinitely long time}\}$. Then $\mathcal{G}_\infty=(\mathcal{V},\mathcal{E}_\infty)$ is connected since $\mathcal
{G}([t,+\infty))$ is connected for all $t\geq0$.

Let $N_i^\infty$ be the neighbor set of node $i$ in graph $\mathcal{G}_\infty$. With Lemma \ref{lem3}, (\ref{s6}) and (\ref{s5}) yield that for any $i=1,\dots,N$ and $j\in N_i^\infty$,
 \begin{equation}\label{s7}
\lim_{t\rightarrow\infty} \langle x_i(t)-{P}_{ X _0}(x_i(t)), x_j(t)-x_i(t)\rangle=0.
\end{equation}

Taking $i_0\in \mathcal{V}$, we define two hyperplanes:
$$
\mathcal{H}_1(t)\doteq \{v|\langle x_{i_0}(t)-{P}_{ X _0}(x_{i_0}(t)), v-x_{i_0}(t)\rangle=0\};
$$
$$
\mathcal{H}_2(t)\doteq \{v|\langle x_{i_0}(t)-{P}_{ X _0}(x_{i_0}(t)), v-{P}_{ X _0}(x_{i_0}(t))\rangle=0\}.
$$

Then $\forall j\in  N_{i_0}^\infty$, (\ref{s7}) implies that
 \begin{equation}
\lim_{t\rightarrow\infty} |x_j(t)|_{\mathcal{H}_1(t)}=0; \quad \lim_{t\rightarrow\infty} |x_j(t)|_{\mathcal{H}_2(t)}=\sqrt{g^\ast}, \nonumber
\end{equation}
which leads to
\begin{equation}\label{s8}
\lim_{t\rightarrow\infty} |{P}_{ X _0}(x_{j}(t))-{P}_{ \mathcal{H}_2(t)}(x_j(t))|=0.
\end{equation}
Because $\mathcal{G}_\infty$ is connected, we can repeat the analysis over the network, then arrive that (\ref{s8}) holds for all $j=1,\dots,N$.

Let $\mathcal {C}_{i_0}(t)=co\{{P}_{ X _{i_0}}(x_{i_0}(t)), {P}_{ X _0}(x_{1}(t)), \dots, {P}_{ X _0}(x_{N}(t)) \}$. Then $\mathcal {C}_{i_0}(t)\subseteq X_{i_0}, \forall t\geq 0$.

Therefore,  with (\ref{s6}) and (\ref{s8}) and according to the structure of $\mathcal{H}_1(t)$ and $\mathcal{H}_2(t)$, there will be a point $z_\ast\in\bigcap_{i_0=1}^N \mathcal {C}_{i_0}(t)\subseteq X_{0} $ for sufficiently large $t$ such that
$$
\langle x_{i_0}(t)-{P}_{ X _0}(x_{i_0}(t)), z_\ast-{P}_{ X _0}(x_{i_0}(t))\rangle>0,
$$
which contradicts (\ref{r9}). Therefore, ${d}^\ast>0$ does not hold, and then the optimal set convergence  follows. \hfill$\square$

\section{Global Consensus}
In this section, we present the consensus analysis.
In order to show the consensus, we have to present a clear estimation of the influence on state agreement by terms $x_i- P _{X_i}(x_i),i=1,\dots, N$.

We first introduce a class of positively invariant set for system (\ref{9}) which characterizes the agreement property in Subsection 5.1. Then the consensus analysis is investigated for directed and bidirectional communication cases, respectively in Subsection 5.2.

\subsection{Invariant Set}
 We define a multi-projection function: $
P_{i_k i_{k-1}\dots i_1}: \mathds{R}^{m}\rightarrow \bigcup_{i=1}^{N} X _i
$ with $i_1,\dots,i_k \in\{1,\dots,N\}, k=1,2,\dots$, by
$$
P_{i_k i_{k-1}\dots i_1}(x)={P}_{ X _{i_k}}{P}_{ X _{i_{k-1}}}\dots {P}_{ X _{i_1}}(x).
$$
Particularly,  $P_{\emptyset}$ is denoted by $P_{\emptyset}(x)=x$ as the case for $k=0$. Let
$$
\Gamma \doteq \{P_{i_k i_{k-1}\dots i_1}| i_1,\dots,i_k \in\{1,\dots,N\},  k=0,1,2,\dots\}
$$
be the set which contains all the multi-projection functions we define.

\begin{figure}
\centerline{\epsfig{figure=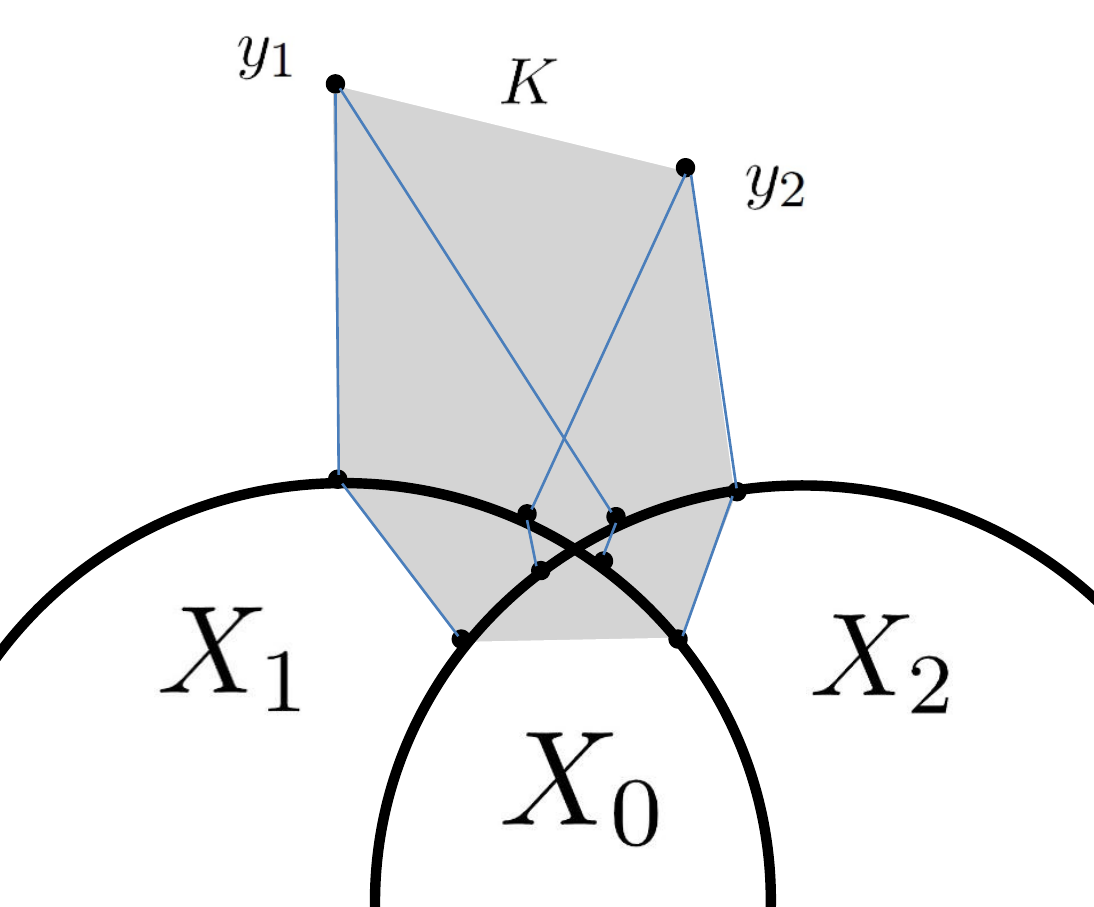, width=0.4\linewidth=0.2}}
\caption{Constructing an invariant set from $K=co\{y_1,y_2\}$.}\label{fig3}
\end{figure}

Furthermore, let $K$ be a convex set in $\mathds{R}^{m}$, and define $\Delta_K$ as
$\Delta_K \doteq co\{P(y)|y\in K,P\in\Gamma\}.$ Denoting $\hat{g}(t)=\max\limits_{i=1,\dots, N}|x_i(t)|_{\Delta_K}^2$, based on a similar analysis as the proof of Lemma \ref{lem2}, it is not hard to find that
$$
D^+\hat{g}(t)\leq 0, \quad t\geq 0.
$$
 This implies,  $\hat{g}(t)\equiv 0$ for all $t\geq t_0$ once we have $\hat{g}(t_0)=0$, which leads to the following conclusion immediately (see Fig. \ref{fig3}).

\begin{lem}\label{lem8}
Let $K$ be a convex set in $\mathds{R}^{m}$. Then $\Delta_K^N\doteq\Delta_K\times\dots\times\Delta_K$ is positively invariant for system (\ref{9}).
\end{lem}

We next establish an important property of the constructed invariant set $\Delta_K^N$.

\begin{lem}\label{lem9}
 $|y|_{K}\leq 2\max_{z\in K}|z|_{ X _{0}}, \;\forall y\in \Delta_K$.
\end{lem}
{\it Proof.} With Lemma \ref{lems1}, any $y\in \Delta_K$ has the following form
$$
y=\sum_{i=1}^{m+1} \lambda_i P^{\langle i \rangle}(z_i),
$$ where $\sum_{i=1}^{m+1} \lambda_i=1$ with $\lambda_i\geq0$,  $P^{\langle i \rangle}\in\Gamma$ and $z_i\in K,\; i=1,\dots,m+1$. Then, by the non-expansiveness property (\ref{r8}), we have that for any $z\in \mathds{R}^m$ and $P_\ast \in \Gamma$,
 \begin{align}
 |{P}_{ X _{0}}(z)-P_\ast(z)|= |P_\ast({P}_{ X _{0}}(z))-P_\ast(z)|\leq |{P}_{ X _{0}}(z)-z|=|z|_{ X _{0}}. \nonumber
 \end{align}
This leads to
\begin{align}
|\sum_{i=1}^{m+1} \lambda_i P^{\langle i \rangle}(z_i)-\sum_{i=1}^{m+1} \lambda_i z_i|&\leq \sum_{i=1}^{m+1} \lambda_i |z_i-P^{\langle i \rangle}(z_i)|\nonumber\\
&\leq  \sum_{i=1}^{m+1}\lambda_i  |z_i-{P}_{ X _{0}}(z_i)|+\sum_{i=1}^{m+1}\lambda_i |{P}_{ X _{0}}(z_i)-P^{\langle i \rangle}(z_i)|\nonumber\\
&\leq2\max_{z\in K}|z|_{ X _{0}}, \nonumber
\end{align}
which implies the conclusion because $\sum_{i=1}^{m+1} \lambda_i z_i\in K$.\hfill$\square$

Now we are ready to reach the global consensus for system (\ref{9}). Let us focus on each coordinate, and denote $x_i^{\ell}(t)$ as the $\ell$-th coordinate of $x_i(t)$. Moreover, let
$$
\phi(t)=\min_{i\in\mathcal {V}}\{x_i^{\ell}(t)\},\quad \varphi(t)=\max_{i\in\mathcal {V}}\{x_i^{\ell}(t)\}
$$
be the minimum and the maximum within all the agents. Denote $H(t)\triangleq\varphi(t)-\phi(t)$. Then a consensus is achieved for system (\ref{9}) if and only if $\lim_{t\rightarrow \infty}H(t)=0$.

In the next subsection, we will prove the global consensus for system (\ref{9}) with directed and bidirectional communications, respectively by showing that $\lim_{t\rightarrow \infty}H(t)=0$.
\subsection{Consensus Analysis}
In this subsection, we propose the consensus analysis.  First we study the directed case.
\begin{prop}\label{thm6}
System (\ref{9}) achieves a global consensus if $\mathcal
{G}_{\sigma(t)}$ is UJSC.
\end{prop}
\noindent {\it Proof.} Based on Proposition \ref{thm5},  we have that $\lim_{t\rightarrow \infty}|x_i(t)|_{ X _{0}}=0$ for all $i=1,\dots,N$. Therefore, for any $\varepsilon>0$, there exists $T_1(\varepsilon)>0$ such that, when $t\geq T_1$,
\begin{equation}
|x_i(t)|_{ X _{0}}\leq\frac{1}{2}\varepsilon, \;i=1,\dots,N
\end{equation}
As a result, according to Lemma \ref{lem9}, for any $y\in\Delta_{co\{x_1(t),\dots,x_N(t)\}}$ with $t>T_1(\varepsilon)$, we have
$$
dist(y,co\{x_1(t),\dots,x_N(t)\})\leq \varepsilon.
$$
Moreover, by Lemma \ref{lem8}, we see that $x_i(\hat{t})\in \Delta_{co\{x_1(t),\dots,x_N(t)\}},i=1,\dots,N$ for all $ t\leq\hat{t}\leq \infty$, which implies that for all $\hat{t}\geq t\geq T_1$, we have
\begin{equation}\label{72}
dist(x_i(\hat{t}),\Delta_{co\{x_1(t),\dots,x_N(t)\}})\leq \varepsilon, \quad i=1,\dots,N.
\end{equation}

We divide the following proof into three steps.

\noindent{\it Step 1:} Take $t_1=T_1$ with $x_{i_0}^{\ell}(t_1)=\phi(t_1)$ and denote $T_0=T+2\tau_D$. In this step, we give bound to $x_{i_0}^{\ell}(t)$ during $t\in[t_1,t_1+(N-1) T_0]$.

Based on (\ref{72}), we see that for all $T_1\leq t<\hat{t}\leq \infty$
\begin{equation}\label{34}
\phi(\hat{t})\geq\phi(t)-\varepsilon;\; \varphi(\hat{t})\leq\varphi(t)+\varepsilon.
\end{equation}

Noting the fact that
\begin{equation}\label{102}
\frac{d}{dt}x_{i_0}^\ell(t)
\leq -(N-1)a^\ast x_{i_0}^\ell(t)+ (N-1)a^\ast(\varphi(t_1)+\varepsilon)+\varepsilon, \;t\geq t_1,
\end{equation}
we obtain
\begin{equation}\label{75}
x_{i_0}^\ell(t)
\leq\mu_1\triangleq \varsigma_0\phi(t_1)+(1-\varsigma_0)\varphi(t_1)+\frac{(N-1)a^\ast+1}{(N-1)a^\ast}\cdot\varepsilon, \; t\in[t_1,t_1+(N-1) T_0].
\end{equation}

\noindent{\it Step 2:} Since $\mathcal
{G}_{\sigma(t)}$ is UJSC,  we can find $i_1\in\mathcal{V}$ and $\tilde{t}_1\geq t_{1}$ such that $(i_0,i_1)\in\mathcal
{G}_{\sigma(t)}$ for $t\in[\tilde{t}_1,\tilde{t}_1+\tau_D)\subseteq [t_{1},t_{1}+T_0)$. In this step, we give bound to $x_{i_1}^{\ell}(t_1)$ during $t\in[\tilde{t}_1+\tau_D,t_1+(N-1) T_0]$.

Similarly to the analysis of (\ref{101}),   when $t\in[\tilde{t}_1,\tilde{t}_1+\tau_D)$, one has
\begin{align}
\frac{d}{dt}x_{i_1}^\ell(t)
&\leq a_\ast(\mu_1-x_{i_1}^\ell(t))+(N-2)a^\ast(\varphi(t_1)+\varepsilon-x_{i_1}^\ell(t))+\varepsilon, \nonumber
\end{align}
which yields
\begin{align}\label{76}
x_{i_1}^\ell(\tilde{t}_1+\tau_D)
&\leq \nu_0(\varphi(t_1)+\varepsilon)+(1-\nu_0)\times
\frac{a_\ast\mu_1+(N-2)a^\ast(\varphi(t_1)+\varepsilon)+\varepsilon}{a_\ast+(N-2)a^\ast}\nonumber\\
&=(1-w_0)\varsigma_0 \phi(t_1)+[1-(1-w_0)\varsigma_0] \varphi(t_1) +\hat{L}_0 \varepsilon\nonumber\\
&\triangleq \theta_1
\end{align}
after some simple manipulations by combining (\ref{75}) and (\ref{76}), where $\hat{L}_0=1+\frac{N}{[a_\ast+(N-2)a^\ast](N-1)}$.

Then, applying (\ref{102}) on node $i_1$ during  $t\in[\tilde{t}_1+\tau_D,t_1+(N-1) T_0]$ will lead to
\begin{align}
x_{i_1}^\ell(t)
&\leq \varsigma_0\theta_1+(1-\varsigma_0)\varphi(t_1)+\frac{(N-1)a^\ast+1}{(N-1)a^\ast}\cdot\varepsilon \nonumber\\
&= m_1 \phi(t_1)+[1-m_1] \varphi(t_1) +L_0 \varepsilon,
\end{align}
for all $t\in[\tilde{t}_1+\tau_D,t_1+N T_0]$, where $m_1=(1-w_0)\varsigma_0^2$ and $L_0=\varsigma_0\hat{L}_0+1+\frac{1}{(N-1)a^\ast}$.

\noindent{\it Step 3:}  We proceed the analysis for $i_2, \dots, i_{N-1}$ with $m_{k}=((1-w_0)\varsigma_0^2)^k, k=2,\dots,N-1$.  Denoting $t_2\triangleq t_1+ (N-1)T_0$, we obtain
\begin{equation}
x_{i_{\varrho}}^\ell(t_2)
\leq m_{N-1} \phi(t_1)+(1-m_{N-1}) \varphi(t_1) +(N-1)L_0 \varepsilon,\;\;\varrho=0,\dots, N-1,
\end{equation}
which implies
\begin{equation}\label{j1}
\varphi(t_2)
\leq m_{N-1} \phi(t_1)+(1-m_{N-1}) \varphi(t_1) +(N-1)L_0 \varepsilon.
\end{equation}
(\ref{34}) and (\ref{j1}) lead to
\begin{align}
H(t_2)
&\leq m_{N-1} \phi(t_1)+(1-m_{N-1}) \varphi(t_1) +(N-1)L_0 \varepsilon-(\phi(t_1)-\varepsilon)\nonumber\\
&= (1-m_{N-1})H(t_1)+[(N-1)L_0+1]\varepsilon
\end{align}

Define a time sequence  $T_1=t_1<t_2<\dots$ with $t_k= t_{k-1}+(N-1)T_0$. Applying the same analysis on each interval $[t_{k-1},t_{k})$ will lead to
\begin{equation}
H(t_{k})
\leq (1-m_{N-1})H(t_{k-1})+[(N-1)L_0+1]\varepsilon, \; k=1,2,\dots.
\end{equation}
As a result, we obtain
\begin{align}\label{79}
H(t_{k+1})&\leq (1-m_{N-1})^{k}H(t_1)+\sum_{j=0}^{k-1}(1-m_{N-1})^j[(N-1)L_0+1]\varepsilon\nonumber\\
&\leq (1-m_{N-1})^{k}H(t_1)+\frac{(N-1)L_0+1}{m_{N-1}}\cdot\varepsilon, \quad k=1,2,\dots
\end{align}

Therefore, noting the fact that $0<m_{N-1}<1$, (\ref{34}) and (\ref{79}) yield
$$
\limsup_{t\rightarrow\infty}H(t)\leq (2+\frac{(N-1)L_0+1}{m_{N-1}})\cdot\varepsilon.
$$
Then $\lim_{t\rightarrow\infty}H(t)=0$ since $\varepsilon$ can be arbitrarily small. This completes the proof. \hfill$\square$

Then the global consensus for bidirectional case is proved by the following conclusion.
\begin{prop}\label{thm3}
System (\ref{9}) achieves a global consensus with bidirectional communications if $\mathcal
{G}_{\sigma(t)}$ is IJC.
\end{prop}
\noindent {\it Proof.} Take $t_1=T_1$ with $x_{i_0}^{\ell}(t_1)=\phi(t_1)$ as the proof of Proposition \ref{thm6}. Then (\ref{72}) and (\ref{34}) still hold.

Denote the first time when $i_0$ has at least one neighbor during $t\geq t_{1}$ as $\tilde{t}_1$, and denote the neighbor set of $i_0$ for $t\in[\tilde{t}_1,\tilde{t}_1+\tau_D)$ as $\mathcal {V}_1$. Next, we show the bound for $i_0$ and $j\in\mathcal {V}_1$ during $t\in[\tilde{t}_1,\tilde{t}_1+\tau_D)$ .

Note that
when $i_0$ has no neighbor during $t\in (t_1,s)$ for $t_1\leq s\leq\infty$, one has that for any $t\in [t_1,s)$,
\begin{equation}\label{92}
|x_{i_0}^\ell(t)-x_{i_0}^\ell(s)|\leq \varepsilon.
\end{equation}  Then, we see that
\begin{align}
x_{i_0}^\ell(t)
&\leq  \hat{\mu}_1 \triangleq 
\hat{\varsigma}_0\phi(t_1)+(1-\hat{\varsigma}_0)\varphi(t_1)+\frac{(N-1)a^\ast+1}{(N-1)a^\ast}\cdot\varepsilon\nonumber
\end{align}
for all $t\in[\tilde{t}_1,\tilde{t}_1+\tau_D]$, where $\hat{\varsigma}_0=e^{-(N-1)a^\ast \tau_D}$

By similar analysis with (\ref{76}), we have that for any $j\in\mathcal {V}_1$,
\begin{equation}\label{36}
x_{j}^\ell(\tilde{t}_1+\tau_D)
\leq\hat{\theta}_1\triangleq\hat{m}_1 \phi(t_1)+(1-\hat{m}_1)\varphi(t_1)  +\hat{L}_0 \varepsilon
\end{equation}
with $\hat{m}_1=(1-w_0)\hat{\varsigma}_0$.

When there is no link between $\mathcal {V}\setminus(\{i_0\}\cup\mathcal {V}_1)$ and $\{i_0\}\cup\mathcal {V}_1$ for  $t\in [\tilde{t}_1+\tau_D,\breve{t})$, applying Lemma \ref{lem8} on the subsystem formed by nodes in $\{i_0\}\cup\mathcal {V}_1$, (\ref{72})  leads to
   \begin{equation}
x_{j}^\ell(t)\leq \hat{\theta}_1+\varepsilon, \; \; t\in [\tilde{t}_1+\tau_D,\breve{t}),\; j\in\{i_0\}\cup\mathcal {V}_1.
\end{equation}

Therefore, defining $\tilde{t_2}$ as the first moment during $t\in[\tilde{t}_1+\tau_D,\infty)$ when there is an edge between $j\in\{i_0\}\cup\mathcal {V}_1$ and $\mathcal {V}\setminus(\{i_0\}\cup\mathcal {V}_1)$, we have
\begin{equation}\label{93}
x_{j}^\ell(t)\leq \hat{\varsigma}_0(\hat{\theta}_1+\varepsilon)+(1-\hat{\varsigma}_0)\varphi(t_1)+\frac{(N-1)a^\ast+1}{(N-1)a^\ast}\cdot\varepsilon
\end{equation}
for $t\in [\tilde{t}_2, \tilde{t}_2+\tau_D]$.

 Denoting $\mathcal {V}_2=\{k\in\mathcal {V}|\mbox{there\ is\ a\ link\ between}\ k\ \mbox{and}\ \{i_0\}\cup\mathcal {V}_1\ \mbox{at}\ \tilde{t}_2\}$, bounds for $x_{k}^\ell(\tilde{t}_2+\tau_D), k\in\mathcal {V}_2$  can be similarly given by
\begin{align}
x_{k}^\ell(\tilde{t}_2+\tau_D)
\leq\hat{m}_2\phi(t_1)+(1-\hat{m}_2)\varphi(t_1)+L_0\varepsilon,
\end{align}
where $\hat{m}_2=((1-w_0)\hat{\varsigma}_0^2)^2$.

Next, $\mathcal {V}_3, \dots, \mathcal {V}_{j_0}$ can be defined until $\mathcal {V}=\{i_0\}\cup\mathcal {V}_1 \cup \dots \cup\mathcal {V}_{j_0}$ since $\mathcal
{G}_{\sigma(t)}$ is JC.  Moreover,  with $\hat{m}_{j_0}=((1-w_0)\hat{\varsigma}_0^2)^{j_0}$, we have
\begin{equation}
x_{i_{\varrho}}^\ell(\tilde{t}_{j_0}+\tau_D)
\leq\hat{m}_{j_0}\phi(t_1)+(1-\hat{m}_{j_0})\varphi(t_1)+L_0(N-1) \varepsilon, \; \varrho=1,\dots, N.
\end{equation}

Therefore, denoting $t_2\triangleq\tilde{t}_{j_0}+\tau_D$, we obtain
\begin{equation}
\varphi(t_2)
\leq\hat{m}_{j_0}\phi(t_1)+(1-\hat{m}_{j_0})\varphi(t_1)+L_0(N-1) \varepsilon,
\end{equation}
which implies
\begin{equation}
H(t_2)\leq(1-\hat{m}_{j_0})H(t_1)+(N-1)L_0\varepsilon.
\end{equation}

Then $\lim_{t\rightarrow\infty}H(t)=0$ holds by similar analysis as the proof of Proposition \ref{thm6}. This completes the proof. \hfill$\square$

With Propositions \ref{thm5}, \ref{thm2}, \ref{thm6} and \ref{thm3}, it is straightforward to see that the main results of the paper, Theorems \ref{thm4} and \ref{thm1} hold.
\section{Conclusions}
This paper addressed an optimal consensus problem
for multi-agent systems.  With jointly connected graphs, the considered multi-agent
system achieved not only consensus, but also optimum by
agreeing within the global solution set of a sum of objective
functions. Assuming that each agent can observe the projection information onto the solution set of its own optimization component and the intersection of all solution sets is nonempty, the original unconstrained optimization problem was converted to an intersection computation problem. Control laws applied to the agents were simple and  distributed. The results showed that a global optimization problem can be solved over a multi-agent network under time-varying communications and limited interactions. Future work includes randomization in the nodes' decision-making and event-based methods in the optimization algorithm design.

\section*{Appendix}

\noindent{\bf A.1\; Proof of Lemma \ref{lem4}}

Based on the definitions of $\theta_i$ and $\eta_i$, when $\theta_i=\eta_i={d}^\ast$ holds for all $i=1,\dots, N$, one has
$$
\lim_{t\rightarrow +\infty}  d_i(t)={d}^\ast, \quad i=1,\dots, N
$$
 Thus, for any $\varepsilon>0$, there exists $T_1(\varepsilon)>0$ such that, when
$t\geq T_1(\varepsilon)$,
\begin{equation}\label{15}
d_i(t)\in [{d}^\ast-\varepsilon, {d}^\ast+\varepsilon],\quad
i=1,\dots,N.
\end{equation}

When ${d}^\ast=0$, then it is easy to see that the conclusion holds because $|x_i(t)|_{ X _i}\leq|x_i(t)|_{ X _0}$ for all $t\geq 0$. Therefore, we just assume ${d}^\ast>0$ in the following.

According to (\ref{14}) and (\ref{12}), it is not hard to find that
\begin{equation}
\frac{d }{dt}d_i(t)\leq -2|x_i|_{ X _i}^2+2\langle x_i- P _{ X _0}(x_i),
\sum_{j \in N_i(\sigma(t))}a_{ij}(x,t)(x_j-x_i)\rangle.
\end{equation}
Furthermore, based on (\ref{15}) and Lemmas \ref{lem3} and \ref{lem2}, one has that when
$t>T_1(\varepsilon)$,
\begin{equation}
\langle x_i- P _{X_0}(x_i),
x_j-x_i\rangle\leq |x_i|_{X_0}\cdot     \left|\ |x_i|_{ X _0}-|x_j|_{X_0}\right|\leq 2\sqrt{{d}(t_0)} \varepsilon
\end{equation}
for all $i=1,\dots,N$ and $j \in N_i(\sigma(t))$.

If the conclusion does not hold, there exist a node $i_0$ and a constant $ M_0>0$ such that
\begin{equation}\label{41}
|x_{i_0}(t_k)|_{ X _{i_0}}=M_0
\end{equation}
for a time serial $0<t_1<\dots<t_k<t_{k+1}<\dots$ with $\lim_{k\rightarrow\infty} t_{k+1}=\infty$. Noting the fact that there is a constant $L>0$ such that  $|a-b|\leq L$ for all $a, b \in \{y|\ |y|^2_{ X _0}\leq {d}(t_0)\}$ since $X_0$ is compact, we have that for all for all $i=1,\dots,N$,
\begin{align}\label{42}
\left|\frac{d}{dt} |x_i(t)|_{ X _i}^2\right| &= \left| 2 \sum_{j \in N_i(\sigma(t))}a_{ij}\langle x_i- P _{ X _i}(x_i),
x_j-x_i\rangle-2|x_i(t)|_{ X _i}^2 \right|\nonumber\\
&\leq 2|x_i(t)|_{ X _0}^2+2 (N-1)a^\ast |x_i(t)|_{ X _0}\cdot |x_j(t)-x_i(t)|\nonumber\\
&\leq 2 {d}(t_0)+2 (N-1)a^\ast \sqrt{{d}(t_0)}L.
\end{align}

Denoting $\tau_0\triangleq\frac{M_0}{2\sqrt{{d}(t_0)+(N-1)a^\ast \sqrt{{d}(t_0)}L}}$ and according to (\ref{41}) and (\ref{42}), we obtain

\begin{equation}
|x_{i_0}(t_k)|^2_{ X _{i_0}}\geq \frac{1}{2}M_0^2, \quad t\in [t_k,t_k+\tau_0],
\end{equation}
which leads to
\begin{equation}
\frac{d }{dt}d_{i_0}(t)\leq -\frac{1}{2}M_0^2+2\sqrt{{d}(t_0)} \varepsilon\leq-\frac{1}{4}M_0^2, \quad t\in [t_k,t_k+\tau_0].
\end{equation}
for all $t_k>T_1$ and $\varepsilon\leq \frac{M_0^2}{8\sqrt{{d}(t_0)}}$. As a result, we have
\begin{equation}\label{43}
d_{i_0}(t_k+\tau_0)\leq {d}(t_0)-\frac{M_0^2\tau_0}{4}+\varepsilon
\end{equation}

Therefore, (\ref{43}) contradicts (\ref{15}) when $\varepsilon<\frac{M_0^2\tau_0}{8}$, which completes the proof. \hfill$\square$

\vspace{2mm}

\noindent{\bf A.2 \; Proof of Lemma \ref{lem10}}

We prove the conclusion by contradiction. Suppose there exists a node $i_0\in\mathcal {V}$ such that $0\leq\theta_{i_0}<\eta_{i_0}\leq {d}^\ast$. Then for any $\varepsilon>0$, there exists $T_1(\varepsilon)>0$ such that, when
$t\geq T_1(\varepsilon)$,
\begin{equation}\label{s4}
{d}_i(t)\in [0, {d}^\ast+\varepsilon],\quad
i=1,\dots,N.
\end{equation}

Take $\zeta_0=\sqrt{\frac{1}{2}(\theta_{i_0}+\eta_{i_0})}$. Then there exists a time serial
$$
0<\hat{t}_1< \dots <\hat{t}_k<\dots
$$
with $\lim_{t\rightarrow\infty}\hat{t}_k=\infty$ such that  $|x_{i_0}(\hat{t}_k)|_{ X _{0}}=\zeta_0$ for all $k=1,2,\dots$.

 According to (\ref{s4}) and Lemma \ref{lem3}, we have that for all $t>\hat{t}_{k_0}$,
\begin{align}
\frac{d }{dt}{d}_{i_0}(t)&\leq 2\sum_{j \in N_{i_0}(\sigma(t))}a_{i_0j}(x,t)\langle x_{i_0}-{P}_{ X _0}(x_{i_0}),
x_j-x_{i_0}\rangle\nonumber\\
&\leq 2(N-1)a^\ast |x_{i_0}(t)|_{ X _0}(\sqrt{{d}^\ast+\varepsilon}-|x_{i_0}(t)|_{ X _0}),\nonumber
\end{align}
which will lead to
\begin{equation}
D^+|x_{i_0}(t)|_{ X _{0}}
\leq -(N-1)a^\ast|x_{i_0}(t)|_{ X _{0}}+(N-1)a^\ast\sqrt{{d}^\ast+\varepsilon}.
\end{equation}
As a result, for $t\in[s,\infty)$ with $s\geq \hat{t}_{k_0}$, we have
\begin{equation}\label{45}
|x_{i_0}(t)|_{X_0}
\leq e^{-(N-1)a^\ast(t-s)}|x_{i_0}(s)|_{X_0}+(1-e^{(N-1)a^\ast(t-s)})\sqrt{{d}^\ast+\varepsilon}.
\end{equation}
We divide the following proof into two cases: directed communications and bidirectional communications.

\noindent{\it Directed Case:} Denote $T_0=T+2\tau_D$. Since $\mathcal
{G}_{\sigma(t)}$ is UJSC, it is not hard to find that there exist $i_1\in \mathcal {V}$ and $\tilde{t}_1$ such that $(i_0,i_1)\in\mathcal
{G}_{\sigma(t)}$ for $t\in[\tilde{t}_1,\tilde{t}_1+\tau_D)\subseteq [\hat{t}_{k_0},\hat{t}_{k_0}+T_0)$.    Then based on (\ref{45}), we obtain
\begin{equation}\label{47}
|x_{i_0}(t)|_{X_0}
\leq \xi_1\triangleq \varsigma_0\zeta_0+(1-\varsigma_0)\sqrt{{d}^\ast+\varepsilon}, \; \;t\in[\hat{t}_{k_0},\hat{t}_{k_0}+(N-1)T_0],
\end{equation}
where $\varsigma_0=e^{-(N-1)^2a^\ast T_0}$. Thus, for $t\in[\tilde{t}_1,\tilde{t}_1+\tau_D)$, one has
\begin{align}
\frac{d }{dt}d_{i_1}(t)&\leq 2[\sum_{j \in N_{i_1}(\sigma(t))\setminus i_0}a_{i_1j}\langle x_{i_1}- P _{X_0}(x_{i_1}),
x_j-x_{i_1}\rangle+a_{i_1 i_0}\langle x_{i_1}- P _{X_0}(x_{i_1}),
x_{i_0}-x_{i_1}\rangle]\nonumber\\
&\leq 2(N-2)a^\ast |x_{i_1}(t)|_{X_0}(\sqrt{{d}^\ast+\varepsilon}-|x_{i_1}(t)|_{X_0})-a_\ast |x_{i_1}(t)|_{X_0}(|x_{i_1}(t)|_{X_0}-\xi_1),
\end{align}
which leads to
\begin{equation}
D^+|x_{i_1}(t)|_{X_0}
\leq -((N-2)a^\ast+a_\ast)|x_{i_1}(t)|_{X_0}+(N-2)a^\ast\sqrt{{d}^\ast+\varepsilon}+a_\ast\xi_1.
\end{equation}
Therefore, we obtain
\begin{equation}
|x_{i_1}(t)|_{X_0}
\leq e^{-((N-2)a^\ast+a_\ast)(t-\tilde{t}_1)}|x_{i_1}(\tilde{t}_1)|_{X_0}+(1-e^{-((N-2)a^\ast+a_\ast)(t-\tilde{t}_1)})\cdot\frac{(N-2)a^\ast\sqrt{{d}^\ast+\varepsilon}+a_\ast\xi_1}{(N-2)a^\ast+a_\ast}\nonumber
\end{equation}
for $t\in[\tilde{t}_1,\tilde{t}_1+\tau_D)$, which implies
\begin{equation}\label{48}
|x_{i_1}(\tilde{t}_1+\tau_D)|_{X_0}
\leq\zeta_1\triangleq w_0\sqrt{{d}^\ast+\varepsilon}+(1-w_0)\xi_1,
\end{equation}
where $w_0$ is defined in (\ref{103}). Furthermore, applying the same analysis of (\ref{45}) on node $i_1$, one has that when $t\in [\tilde{t}_1+\tau_D,\infty)$,
\begin{equation}\label{49}
|x_{i_1}(t)|_{X_0}
\leq e^{-(N-1)a^\ast(t-(\tilde{t}_1+\tau_D))}\zeta_1+(1-e^{-(N-1)a^\ast(t-(\tilde{t}_1+\tau_D))})\sqrt{{d}^\ast+\varepsilon},
\end{equation}
Combing (\ref{47}), (\ref{48}) and (\ref{49}), we obtain
\begin{equation}\label{50}
|x_{i_1}(t)|_{X_0}\leq m_1 \zeta_0+(1-m_1)\sqrt{{d}^\ast+\varepsilon}, \quad
\end{equation}
for all $ t\in[\tilde{t}_1+\tau_D,\hat{t}_{k_0}+(N-1)T_0]$, where $m_1=(1-w_0)\varsigma_0^2$. (\ref{50}) also holds for $i_0$ since $0<\varsigma_0<m_1<1$.

We can proceed to find a node $i_2\in \mathcal {V}$ such that there is an arc leaving from $\{i_0, i_1\}$  entering $i_2$ in $\mathcal
{G}([\hat{t}_{k_0}+T_0,\hat{t}_{k_0}+2T_0))$ because $\mathcal
{G}_{\sigma(t)}$ is uniformly jointly strongly connected. Meanwhile, similar analysis will result in estimations for agent $i_2$ with the form (\ref{50}) by $m_2=((1-w_0)\varsigma_0^2)^2$.

Repeating similar analysis on time intervals $[\hat{t}_{k_0}+2T_0,\hat{t}_{k_0}+3T_0],\dots, [\hat{t}_{k_0}+(N-2)T_0,\hat{t}_{k_0}+(N-1)T_0]$ respectively, and finally, by $m_{N-1}=((1-w_0)\varsigma_0^2)^{N-1}$, we obtain
\begin{equation}
|x_{i}(\hat{t}_{k_0}+NT_0)|_{X_0}
\leq m_{N-1}\zeta_0+(1-m_{N-1})\sqrt{{d}^\ast+\varepsilon}, \; i=1,\dots,N,
\end{equation}
which yields
\begin{equation}\label{51}
d(\hat{t}_{k_0}+NT_0)
\leq m_{N-1}\zeta_0+(1-m_{N-1})\sqrt{{d}^\ast+\varepsilon}.
\end{equation}

Note that, (\ref{51}) contradicts the definition of ${d}^\ast$ since $m_{N-1}\zeta_0+(1-m_{N-1})\sqrt{{d}^\ast+\varepsilon}<\sqrt{{d}^\ast}$ for sufficiently small $\varepsilon$. The conclusion holds.

\noindent{\it Bidirectional Case:} When $i_0$ has no neighbor for $t\in [\hat{t}_{k_0},s]$, by (\ref{12}) we see that
 \begin{equation}
|x_{i_0}(t)|_{ X _{0}}\leq |x_{i_0}(\hat{t}_{k_0})|_{ X _{0}}=\zeta_0, \; t\in [\hat{t}_{k_0},s].
\end{equation}

Denote the first moment when $i_0$ has at least one neighbor during $t\in [\hat{t}_{k_0},\infty)$ as $\tilde{t}_1$, and denote the neighbor set of $i_0$ for $t\in[\tilde{t}_1,\tilde{t}_1+\tau_D)$ as $\mathcal {V}_1$. Then,  by a similar analysis as (\ref{47}), one has
\begin{equation}
|x_{i_0}(t)|_{ X _{0}}
\leq \hat{\xi}_1\triangleq \hat{\varsigma}_0\zeta_0+(1-\hat{\varsigma}_0)\sqrt{{d}^\ast+\varepsilon},\,\; t\in[\tilde{t}_1,\tilde{t}_1+\tau_D]
\end{equation}
with $\hat{\varsigma}_0=e^{-(N-1)a^\ast \tau_D}$. Thus, according to the same process by which we obtain (\ref{48}), one also obtains
\begin{equation}
|x_{i_1}(\tilde{t}_1+\tau_D)|_{ X _{0}}
\leq w_0\sqrt{{d}^\ast+\varepsilon}+(1-w_0)\hat{\xi}_1=\hat{m}_1\zeta_0+(1-\hat{m}_1)\sqrt{{d}^\ast+\varepsilon},
\end{equation}
where $\hat{m}_1=\hat{\varsigma}_0(1-w_0)$.

 Similarly, we can define $\tilde{t}_2$ as the first moment when there is another node connected to $\{i_0\}\cup\mathcal {V}_1$ during $t\geq\tilde{t}_1+\tau_D$.  Let $\mathcal {V}_2$ be the node set which connect to $\{i_0\}\cup\mathcal {V}_1$ at $\tilde{t}_2$. Since we have the dwell time for $\sigma(t)$, without loss of generality, we can always assume that all the links between $\{i_0\}\cup\mathcal {V}_1$ and $\mathcal {V}_2$ last for at least $\tau_D$ time starting from $\tilde{t}_2$. Moreover, similar estimations will lead to
\begin{equation}
|x_{i_2}(\tilde{t}_2+\tau_D)|_{ X _{0}}\leq \hat{m}_2\zeta_0+(1-\hat{m}_2)\sqrt{{d}^\ast+\varepsilon} \nonumber
\end{equation}
for all $i_2\in \{i_0\}\cup\mathcal {V}_1\cup\mathcal {V}_2$, where $\hat{m}_2=(\hat{\varsigma}_0(1-w_0))^2$.

Furthermore, since $\mathcal
{G}_{\sigma(t)}$ is JC, we can always proceed the upper process  until $\mathcal {V}=\{i_0\}\cup\mathcal {V}_1 \cup \dots \cup\mathcal {V}_{j_0}$, and then we obtain
\begin{eqnarray}
|x_{i}(\tilde{t}_{j_0}+\tau_D)|_{ X _{0}}
\leq \hat{m}_{j_0}\zeta_0+(1-\hat{m}_{j_0})\sqrt{{d}^\ast+\varepsilon},\nonumber
\end{eqnarray}
with $\hat{m}_{j_0}=(\hat{\varsigma}_0(1-w_0))^{j_0}$, which  contradicts the definition of ${d}^\ast$.  Then the conclusion holds for bidirectional case.

The proof is completed. \hfill$\square$

\end{document}